\documentclass[prl,aps,amssymb,amsmath,twocolumn,showpacs]{revtex4}
\usepackage{graphicx,epsfig}
\usepackage{psfrag}
\usepackage{dcolumn}
\newcommand{\eq}[1]{Eq.~{(\ref{#1})}}
\newcommand{\fig}[1]{Fig.~{\ref{#1}}}

\newcommand{\sample}[1]{\overline{#1}}

\begin{document}

\title{Universal statistics of the critical depinning force of elastic systems
in random media}

\author{C.~J.~Bolech} 
\author{Alberto Rosso} 

\affiliation{Universit\'e de Gen\`eve, DPMC, 24 Quai Ernest Ansermet, CH-1211
Gen\`eve 4, Switzerland}

\date{February $2^{nd}$, 2004}

\begin{abstract}
We study the rescaled probability distribution of the critical depinning force
of an elastic system in a random medium. We put in evidence the underlying
connection between the critical properties of the depinning transition and the
extreme value statistics of correlated variables. The distribution is Gaussian
for all periodic systems, while in the case of random manifolds there exists a
family of universal functions ranging from the Gaussian to the Gumbel
distribution. Both of these scenarios are a priori experimentally accessible
in finite, macroscopic, disordered elastic systems.
\end{abstract}


\maketitle

%\section{Introduction}

The study of disordered elastic objects sheds light on the physics of numerous
experimental systems, including domain walls in magnetic
\cite{lemerle_domainwall_creep} or ferroelectric \cite{tybell_ferro_creep}
materials, contact lines of liquid menisci on a rough substrate
\cite{rosso_width2} and propagating cracks in solids
\cite{bouchaud_bouchaud}. All these are accurately modeled as elastic
manifolds in the presence of randomness. A second class of disordered elastic
systems is given by periodic structures such as charge density waves
\cite{gruner_revue_cdw}, vortex flux lines in type-II superconductors
\cite{blatter_vortex_review} and Wigner crystals \cite{chitra_wigner_long}. It
has been proven that the periodic nature of these structures carries to the
same universal behavior as elastic interfaces in the presence of periodic
disorder \cite{middleton_cdw,chauve_2loop}.

A continuing challenge is to understand the response of these systems to an
applied external drive per unit surface, $f$. Two regimes are present at zero
temperature: whenever $f$ is smaller than certain critical threshold $f_c$ the
interface is pinned, but beyond this point ($f>f_c$) the system undergoes a
{\em depinning} transition
\cite{nattermann_stepanow_depinning,narayan_fisher_depinning,narayan_fisher_cdw,chauve_2loop,2loop}
and continues to move with a non-zero average velocity. Throughout all of the
low temperature regime the dynamics is strongly influenced by the presence of
this threshold. It is thus interesting to characterize $f_c$ in detail, and in
particular, to study its sample-dependent fluctuations. We remark that the
critical force is actually the biggest one that still pins the system. New
insights shall then be expected from exploring the eventual connections with
the theory of extreme value statistics.

In a way reminiscent to the central limit theorem for the sum of uncorrelated
variables, there exists a fundamental result for the statistics of extremes
\cite{galambos,coles}. It states that only three limits for the maxima of
independent identically distributed (iid) variables are possible: the Weibull
family (for upper bounded variables), the Fr{\'e}chet family (for
power-law-tailed variables) or the Gumbel distribution (for exponential-tailed
variables). However, due to the complicated dynamics of elastic systems, we
expect that strong correlations should play a crucial role in determining the
limiting distribution of its thresholds. Indeed, the depinning transition is
an example of a critical phenomenon with well-determined universal exponents,
which implies strong, long ranged, well-characterized correlations. The few
results known related to extreme values of correlated variables concern short
range or weak correlations
\cite{coles,bouchaud_mezard,carpentier_extreme,monthus_ledu}. The opposite
case has only been started to be discussed
\cite{dean_extreme,amaral_extreme,racz_extreme}. Here, we identify and study
an example of extremes of strongly correlated variables with clear physical
interest.

To pose the problem, let us consider a fixed disorder realization on a finite
system. To every configuration of the elastic interface, ${\{\alpha\}}$, we
associate a depinning force, $f_d^{\{\alpha\}}$, given by the smallest
non-negative force so that at least one point of the interface has instant
non-negative velocity \cite{middleton_theorem, rossokrauthI}. The critical
force of the realization follows naturally:
\begin{equation}
  f_c^r = \max_{\{\alpha\}} \; \{f_d^{\{\alpha\}}\} \; .
\end{equation}
In this letter we will concentrate on the statistics of $f_c^r$. We shall
identify the universal aspects of its distribution and stress the important
differences between the periodic and random manifold cases. We shall find that
whilst for periodic systems the distribution is always a Gaussian, for random
manifolds it belongs to a family that interpolates between the Gaussian and
the Gumbel\footnote{We recall that the Gumbel distribution is given by
$(\pi/\sqrt{6})g(x)\exp(-g(x))$, where $g(x)=\exp(-\pi x/\sqrt{6}-\gamma)$ and
$\gamma$ is Euler's constant.}.

%\section{Model}

The focus of our attention will be a $d$-dimensional interface propagating
transversally on a $(d+1)$-dimen\-sional space; for clarity we describe now
the case of a line ($d=1$). At any given time, the line defines a
single-valued function $h(x)$, where $x$ is the longitudinal coordinate. Its
zero temperature overdamped dynamics is governed by the following equation of
motion,
\begin{equation} 
\label{e:continuum_motion}
\partial_t h(x,t) = - \frac{\partial E(\left\{h,x \right\})}{\partial
h(x,t) } = c \nabla^2 h + f + \eta(x,h (x,t)) \ .
\end{equation}
The functional $E(\left\{h,x \right\})$ represents the total energy, including
the harmonic elastic energy and the potential energy terms due to the drive
$f$ and the short-range correlated disorder force $\eta(x,h)$. This is a
non-linear equation whose analytic solution is unknown.  As alluded to in the
beginning, properties of the disorder distinguish two different cases. If
$\eta(x,h)$ is a periodic function, \eq{e:continuum_motion} describes a
periodic system, whereas the random manifolds are characterized by a
non-periodic disorder.

\begin{figure}[b]
\psfrag{xr}[][]{\small $\phantom{mmm}M=kL^{\zeta}$}
\psfrag{xp}[][]{\small $\phantom{mmmmm}M=\mathrm{const.}$}
\psfrag{L}[][]{\small $L$}
\includegraphics[width=\columnwidth]{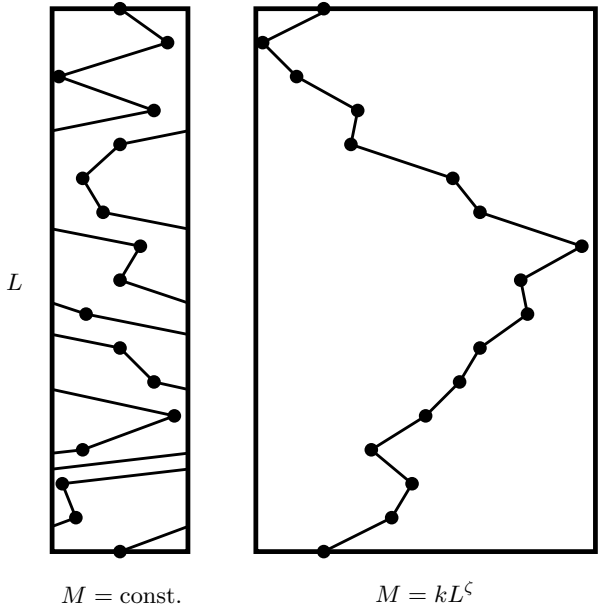}
\caption{\label{f:model} Elastic lines in a disordered medium. The left schema
illustrates a periodic system whereas the right one shows a random manifold.}
\end{figure}

Our approach to studying \eq{e:continuum_motion} will be numerical. In
particular, for each disorder realization, we find the critical force $f_c^r$
using an exact algorithm \cite{rosso_krauth_longrange} that locates the
critical line $h_c^r(x)$ ({\em i.e.} the ultimate pinned configuration).
\fig{f:model} schematizes the model: we keep $h$ as a continuous variable and
discretize the coordinate $x\to i$; periodic boundary conditions are used so
that $h$ has period $M$ and $i$ has period $L$
\cite{rosso_krauth_longrange,rosso_krauth_manifold}. For a periodic disorder
$M$ is kept constant as $L$ goes to infinity. In this case the interface winds
around the system and feels the disorder periodicity. For a random manifold
both $L$ and $M$ go to infinity. The most physical scaling corresponds to $M
\sim L^\zeta$, where $\zeta$ is the roughness exponent at the depinning
transition, and allows the transversal dimension to scale with the lateral
extension of the interface \cite{narayan_fisher_depinning,chauve_2loop}.

%As every critical transition depinning has associated a divergent correlation
%length, $\xi$, that above threshold is related to the velocity-velocity
%correlation. Past studies have found that
%\cite{nattermann_stepanow_depinning},
%\begin{align}
%\label{e:nu}
%\xi \sim |f-f_c|^{-\nu}& &\mathrm{with}& &\nu = \frac{1}{2-\zeta} \; .
%\end{align}

A general theorem for disorder-controlled transitions, shows that on a
finite-volume ($L^{d+\zeta}$), any divergent correlation length scales with an
exponent \cite{chayes,narayan_fisher_depinning,chauve_2loop}
\begin{equation}
\label{e:nuFSbound}
\nu_{\mathrm{FS}} \ge \frac{2}{d+\zeta} \; .
\end{equation}
The identification between $\nu_{\mathrm{FS}}$ and the critical correlation
length exponent $\nu$ is a very delicate problem that has been discussed in
the literature \cite{zimanyi}. In any case, for elastic systems it was
determined that
\cite{narayan_fisher_depinning,narayan_fisher_cdw,2loop,middleton_cdw},
\begin{equation}
\label{e:nuFS}
\sigma_{f_c}^2\equiv\sample{|f_c^r-\sample{f_c^r}|^2} \sim
L^{-\frac{2}{\nu_{\mathrm{FS}}}} \; ,
\end{equation}
where the over-bar indicates disorder average.

%\section{Periodic System}

\begin{figure}[t]
\psfrag{x}[][]{\small $\hat{f}_c$}
\psfrag{y}[][]{\small $p(\hat{f}_c)$}
\psfrag{xxx}[][]{\scriptsize $\ell$}
\psfrag{yyy}[][]{\scriptsize $|\mathrm{cov}(\ell)/\mathrm{cov}(0)|$}
\includegraphics[width=\columnwidth]{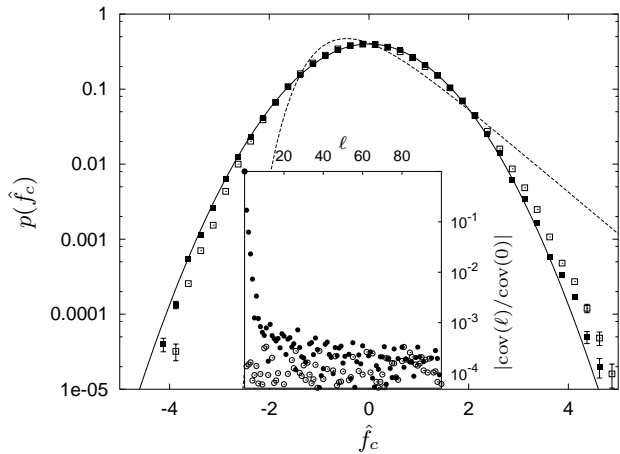}
\caption{\label{f:gaussian} Rescaled distributions for a periodic system. The
squares correspond to the results for a periodic system ($M=8$) with $L=64$
(open) and $L=1024$ (filled); undisplayed error bars are smaller than
the symbol size. The lines are the Normal (solid) and Gumbel
(dashed) distributions. Convergence with $L$ towards the Gaussian limit is
evident. The values for the skewness ($\gamma_1<0.03$) and kurtosis
($\gamma_2<0.01$) are consistent with this limit. Inset: covariance of the
$\eta(i,h_c^r)$'s (filled) compared with that of a set of random
numbers (open).}
\end{figure}

Let us start discussing the periodic case, where by definition $\zeta=0$. We
have performed multiple simulations for $d=1$, determining the critical force
for different disorder realizations. We verify that the scaling relation given
in \eq{e:nuFS} is obeyed with $\nu_{\mathrm{FS}} \simeq 2$ (in agreement with
the results of previous simulations \cite{middleton_cdw}). In order to
highlight the universal behavior in the thermodynamic limit, it is convenient
to rescale the forces and compare always zero-mean unit-variance
distributions, $p(\hat{f}_c)$ with
$\hat{f}_c=(f_c^r-\sample{f_c^r})/\sigma_{f_c}$. In \fig{f:gaussian} we show
that the limit distribution of $p(\hat{f}_c)$ is Gaussian. Finite size effects
are very small and the limiting behavior is robust against changes in the
nature of the disorder, the period $M$, the boundary conditions or any
microscopic parameter like the elastic constant (in all figures, the number of
samples ranges in $10^5$-$10^6$). Simulations done for $d=2$ also confirm this
scenario. The generic nature of this result suggests that, in all periodic
systems, the limiting distribution of thresholds (critical field, current,
etc.) is Gaussian.

To understand this result, we study the correlation of the critical
disorder-forces, $\eta(i,h_c^r)$, defined as the pinning forces on the
different points of the critical line. Since the configuration is static, the
critical force is obtained as minus the average pinning force.
%,
%\begin{equation}
%  f^r_c = - \frac{1}{L^d}\sum_{i} \eta(i,h_c^r) \; .
%\end{equation}
The degree of correlation among the $\eta(i,h_c^r)$'s plays thus a defining
role in determining the final thermodynamic distribution of critical
forces. Evidently, if they were uncorrelated, the distribution would surely be
Gaussian, on the general grounds of the central limit theorem. We studied
therefore their covariance, defined as
\begin{equation}
  \mathrm{cov}(\ell) = \frac{1}{L}\sum_{i}
  \sample{(\eta(i+\ell,h_c^r)+\sample{f^r_c}) (\eta(i,h_c^r)+\sample{f^r_c})}
  \; .
\end{equation}
We found that, in the case of fixed finite $M$, the correlation decreases very
fast (at least exponentially) and the variables quickly become uncorrelated
(see the inset in \fig{f:gaussian}),
explaining the observed convergence of $p(\hat{f}_c)$ towards a Gaussian
distribution. The $\eta(i,h_c^r)$'s self-average in the
thermodynamic limit, in turn implying that $\nu_{\mathrm{FS}}=2/d$,
saturating the bound of \eq{e:nuFSbound}.

%\section{Random Manifold}

\begin{figure}[t]
\psfrag{x}[][]{\small $\hat{f}_c$}
\psfrag{y}[][]{\small $p(\hat{f}_c)$}
\includegraphics[width=\columnwidth]{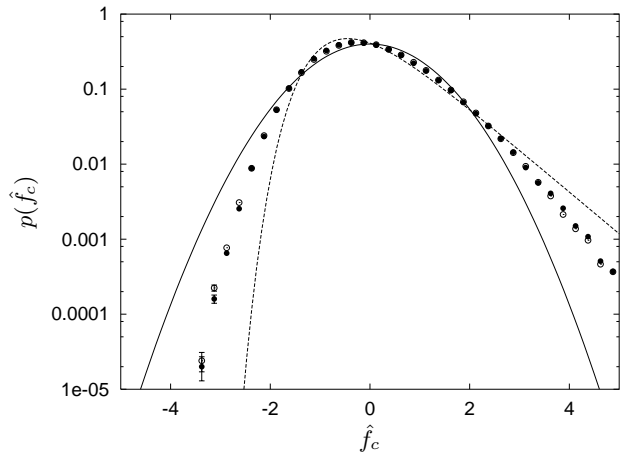}
\caption{\label{f:manifold} Rescaled distributions for a random manifold. The
circles correspond to the results for a scaling with $M=L^{\zeta}$ for $L=64$
(open) and $L=768$ (filled). The lines and error bars are as in
\fig{f:gaussian}.  Convergence with $L$ towards a non-trivial limit is
evident. The extrapolated values for the skewness and kurtosis are $\gamma_1
\to 0.57\pm0.01$ and $\gamma_2 \to 0.63\pm 0.03$.}
\end{figure}

We direct now our attention to the case of random manifolds. We find
$\nu_{\mathrm{FS}}=1.33 \pm 0.01$, verifying for $d=1$ the scaling relation
$\nu_{\mathrm{FS}}=\nu=(2-\zeta)^{-1}$
\cite{nattermann_stepanow_depinning,narayan_fisher_depinning,2loop}, with
$\zeta \sim 1.26 \pm 0.01$ \cite{leschhorn_tang,rosso_krauth_manifold}. We
stress that this relation is valid, as will be clearly shown later, only in
correspondence to the physical scaling $M \sim L^\zeta$
(cf.~Ref.~[\onlinecite{ramanathan}]). In this case the distribution is
non-trivial; our results for a $d=1$ random manifold are presented in
\fig{f:manifold}. We find a distribution that converges as $L\to \infty$ to a
universal curve that clearly deviates from the Gaussian in the direction of
the Gumbel. Previous studies on cellular automaton models have found that
distributions of near-threshold depinning forces are also universal
\cite{vandembroucq1,vandembroucq2}.

The proper scaling for the random manifold case should be defined with
care. We consider $M=kL^\zeta$ with $k$ a constant independent of $L$. For
each fixed $k$, we find that the distribution of critical forces has a
non-trivial thermodynamic limit. In particular, for $k\to0$ this limit tends
to a Gaussian, as seen in the periodic case, and in the opposite limit
($k\to\infty$) the distribution approaches Gumbel's. We verified this behavior
for $d=1$ and results for $d=2$ confirm the same scenario. The most natural
definition of a random manifold system corresponds to considering values
around certain, model dependent, $k_w=w/L^\zeta$ (where $w$ is the average
width of the elastic line \cite{rosso_width}). The ratio $k/k_w$ then
parameterizes a universal family of functions that crosses over from the
Gaussian to the Gumbel distribution.  This set of extremal distributions is
universal in the sense of being independent of all microscopic parameters of
the model and fixed solely by $d$ and the ratio $k/k_w$.

\begin{figure}[t]
\psfrag{x}[][]{\small $f_c$}
\psfrag{y}[][]{\small $H_M(f_c)$}
\includegraphics[width=\columnwidth]{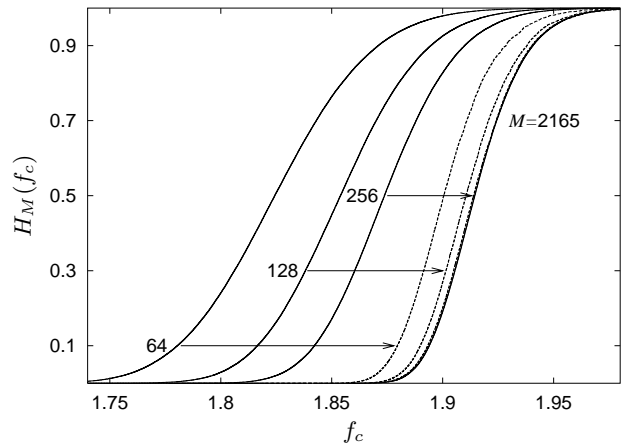}
\caption{\label{f:H} Cumulative distributions of critical forces of lines with
$L=256$. Different solid lines correspond to different $M$'s as indicated. The
average width is $w\sim100$ ($k_w\sim0.1$). The rescaling according to
\eq{e:H} of the curves for $M=64,128,256$ to $M=2165$ ($k=2$), is shown with
dashed lines.}
\end{figure}

To understand the {\it Gumbel limit}, let us start from the problem of iid
variables. Given a set of $n$ variables with $Z_n$ its maximum, the
associated cumulative probability distribution, $H_n(z)=P(z>Z_n)$, obeys:
\begin{equation}
\label{e:H}
  H_{n}(z)= \left[ H_m(z) \right]^{\frac{n}{m}} \; .
\end{equation}
This is the crucial property that allows to prove the theorem on the
statistics of extreme values mentioned at the beginning. We can now go back to
the case of random manifolds. When the aspect ratio of the samples becomes
very wide, $k \gg k_w$, the system can be thought to be split into $k/k_w$
approximately independent regions. This is because the configurations with
non-zero depinning force proliferate and their correlation decreases. In this
case \eq{e:H} is asymptotically verified and the results of the theorem are
still valid as shown by the statistics of extremes of {\it stationary}
sequences \cite{galambos,coles}. In order to illustrate this point, we study
explicitly the cumulative probability of critical forces, $H_M(f_c)$. We
present in \fig{f:H}, curves corresponding to different values of $M$ for $L$
large enough that the finite size effects are indistinguishable in the scale
of the figure. We used \eq{e:H} to scale the distributions towards one
corresponding to a much larger $M$. Let us separate our discussion in three
regimes: $k<k_w$, $k\agt k_w$ and $k\gg k_w$. The figure shows that whereas in
the first case the scaling of \eq{e:H} is violated, in the other two it is
obeyed with increasingly better accuracy as $k$ grows. This demonstrates that,
in the third case, the critical force can be thought of as the maximum among
forces sampled from a distribution like that in \fig{f:manifold}. Since the
tail behavior of the latter is bounded between Gumbel and Gaussian (both
belonging to the {\it domain of attraction} of the Gumbel distribution
\cite{galambos}), it follows that in the $k\to\infty$ limit the distribution
should be Gumbel, as opposed to Fr{\'e}chet or Weibull.

Let us discuss briefly the implications for other scalings of $M$. At this
stage it should be clear that, for any other scaling, the distribution
converges to a non-parametric thermodynamic limit. Consider, for the sake of
concreteness, $M\sim L^{\zeta'}$: on the one hand, for $\zeta'>\zeta$, the
convergence is towards the Gumbel distribution; on the other hand, for
$\zeta'<\zeta$, the limit is the Gaussian.

%\section{Summary}

As we have seen, the study of the distributions of maximum thresholds unravels
a very rich and complex scenario. For all periodic systems the distribution is
a Gaussian, --a non-trivial result from the viewpoint of the theory of extreme
value statistics. This result rests on the asymptotic validity of the central
limit for the critical depinning forces and implies
$\nu_{\mathrm{FS}}=2/d$. For random manifolds the situation is more intricate
and there exists a universal family that interpolates continuously from the
Gaussian to the Gumbel distribution. To our knowledge, this is the first time
a set of distributions of extremes with these characteristics was identified,
(there is another recent example of a family that interpolates between these
two limits, but with no interpretation in terms of extremes
\cite{antal.1overf.01}). The criticality of the system is crucial to obtain
such varied behaviors. Indeed, it is the infinite range of the correlations
that build up along the critical manifold that is responsible for the strong
deviations from the standard theory for iid variables.

It would be extremely interesting to verify these scenarios experimentally. We
believe, good candidate measurements in periodic systems would be those of
superconducting critical currents, or charge density waves. For propagating
interfaces the usual set-ups involve $M \ggg w$. To build a distribution in
this case one should be able to measure the critical thresholds for a given
tiling of the sample, then compare distributions among different tiling
widths. On a different line, we already started to look into the analytical
characterization of the family of distributions that we found for the random
manifold case. This might provide deep insights into the theory of extremes of
correlated variables that is still in its infancy.

%\vskip 2 mm
%We thank ...

We acknowledge the illuminating discussions with M.~Droz, P.~Le~Doussal,
W.~Krauth, M.~M\'ezard, Z.~R\`acz and K.~J.~Wiese. We thank T.~Giamarchi and
W.~Krauth for their comments on the manuscript.

%\bibliographystyle{apsrev}
%\bibliography{totphys,CDW}

\end{document}